\documentclass[%
 reprint,
 amsmath,amssymb,
]{revtex4-2}

\usepackage{graphicx}
\usepackage{dcolumn}
\usepackage{bm}
\usepackage{ulem}

\usepackage{amssymb} 
\usepackage{multirow}
\usepackage{booktabs}
\usepackage{color}     
\usepackage{amsmath}
\usepackage{multirow}  

\usepackage{hyperref} 
\hypersetup{
            colorlinks=true,
            linkcolor=blue,
            filecolor=magenta,      
            urlcolor=cyan, 
            citecolor=red,}
            
\begin{document}

\preprint{APS/123-QED}

\title{Anomaly in first-forbidden transitions of $^{176}$Lu} 

\author{Jing-Wen Ran}%
\affiliation{School of Physical Science and Technology, Southwest University, Chongqing 400715, China}%

\author{Long-Jun Wang}
\email{longjun@swu.edu.cn}
\affiliation{School of Physical Science and Technology, Southwest University, Chongqing 400715, China} 

\date{\today}

\begin{abstract}
  $^{176}$Lu is the key nucleus for understanding the evolution of planetary bodies in the solar system, and the temperature and nucleosynthesis of the $s$ process, due to its very long terrestrial half-life $T_{1/2} \approx 3.7\times 10^{10}$ years. The very long half-life is caused by two anomalous first-forbidden transitions from the $7^-$ state with extremely large comparative half-life log$ft \approx 19$ which have never been appeared in other cases of the existing nuclear databases. We analyze the underlying mechanism and reason for the anomaly in first-forbidden transitions of $^{176}$Lu for the first time, which is based on the projected shell model. It is found that the possible $K$-forbidden nature is indispensable for describing the two extremely weak first-forbidden transitions, and the transition strengths are very sensitive to the detailed configuration mixing and $K$ mixing in the nuclear wave functions. The half-life of the $7^-$ state is calculated to be $T_{1/2} \approx 1.95 \times 10^{10}$ years.
\end{abstract}

\maketitle



\begin{figure}
\begin{center}
  \includegraphics[width=0.48\textwidth]{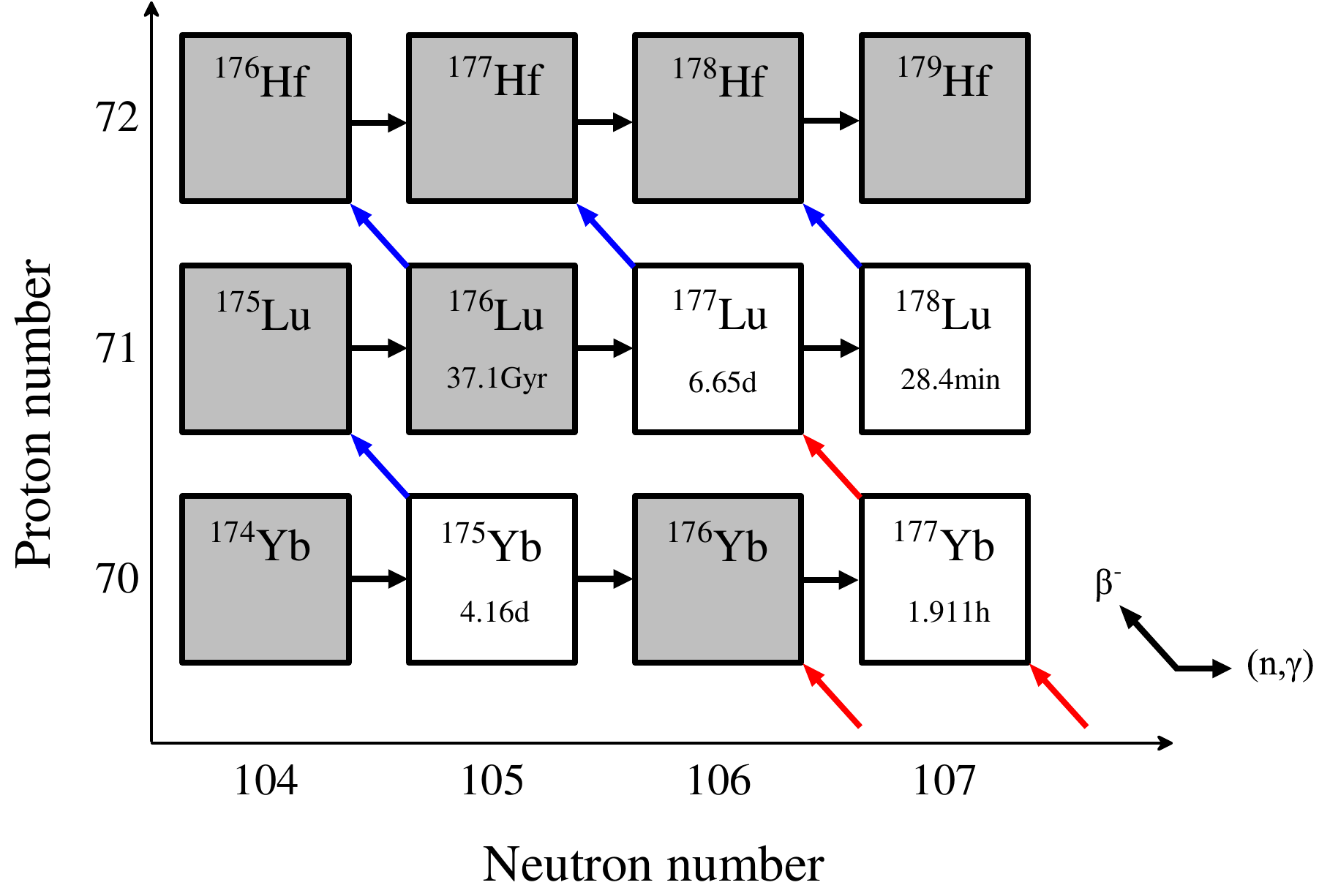}
  \caption{\label{fig:nuclear_chart} (Color online) An excerpt of the chart of nuclides near the $s$-process branching point at $^{176}$Lu. Stable and primordial (other unstable) nuclides are shown in grey (open) squares. $s$- ($r$-) process $\beta^-$ decay is labeled by blue (red) arrows. $^{176}$Lu and $^{176}$Hf are $s$-only-process nuclides which are shielded from the $r$-process by the isobaric nuclide $^{176}$Yb.  } 
\end{center}
\end{figure}

\begin{figure}
\begin{center}
  \includegraphics[width=0.49\textwidth]{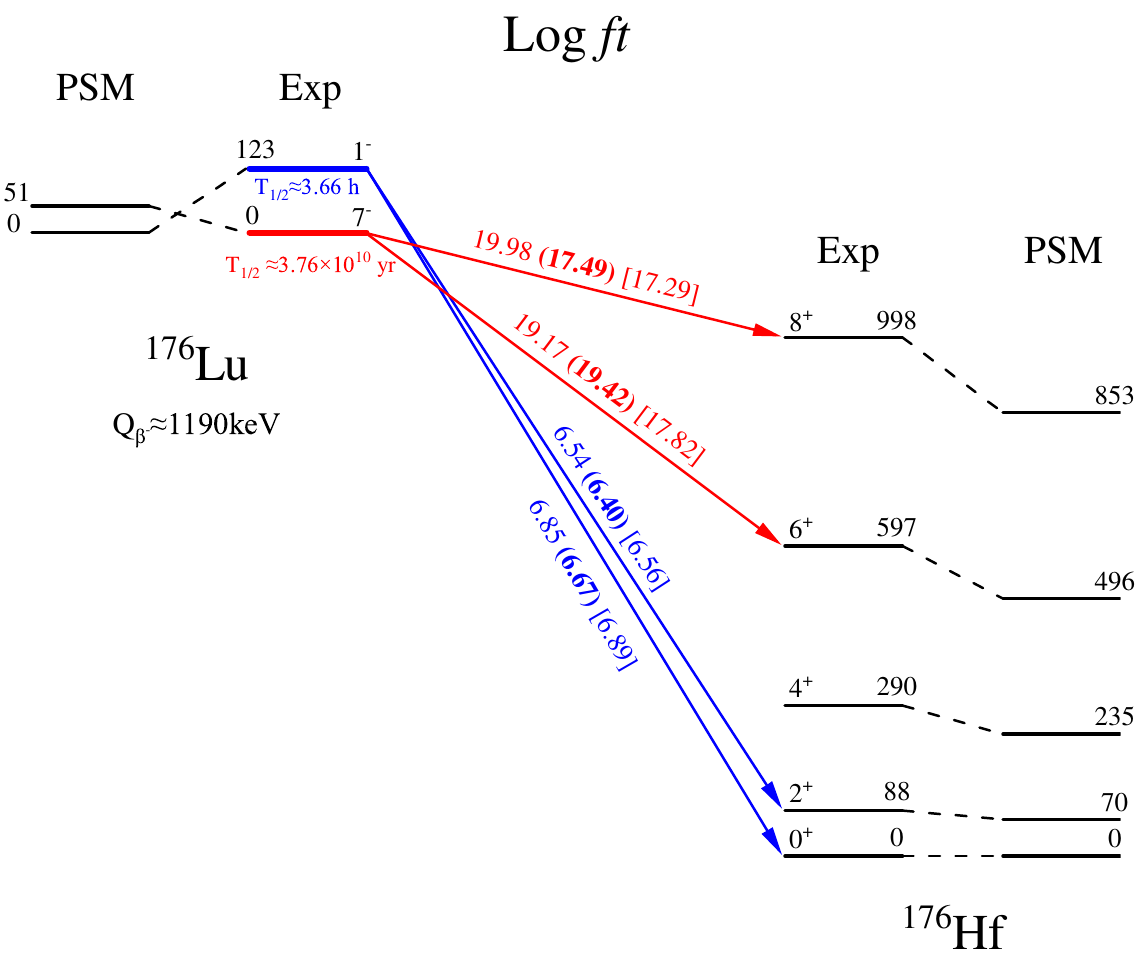}
  \caption{\label{fig:decay_scheme} (Color online) The $\beta$-decay scheme of the $7^-$ ground state and $1^-$ isomer of $^{176}$Lu. The PSM calculations for the spin-parity and excitation energy (in keV) of the low-lying levels are compared with the data evaluated by NNDC \cite{NNDC}. The calculated Log$ft$ values with large (small) configuration space are shown in parenthesis (in bracket) and compared with the evaluated values adopted by NNDC \cite{NNDC}. See the text for details.   } 
\end{center}
\end{figure}

$^{176}$Lu is a mysterious nuclide which is a very long-lived primordial isotope with a half-life $T_{1/2} \approx 3.7\times 10^{10}$ years, although the corresponding $\beta$-decay $Q$ value is not negligible with $Q_{\beta} \approx 1.2$ MeV (see Figs. \ref{fig:nuclear_chart} and \ref{fig:decay_scheme}) \cite{NNDC, livechart, Nucl_Data_Sheet_for176_2006}. The reason is due to the fact that the ground state of $^{176}$Lu can only $\beta^-$ decay to the daughter nucleus $^{176}$Hf by two anomalous first-forbidden transitions, which are extremely weak with large comparative half-life log$ft \approx 19$. Actually, they are the only two strange first-forbidden transitions with log$ft \gg 12$ in the known evaluated nuclear databases \cite{NNDC, livechart}. More interestingly, $^{176}$Lu has a very low-lying isomer, with spin-parity $1^-$ and excitation energy about 120 keV, which can $\beta^-$ decay to $^{176}$Hf by much stronger first-forbidden transitions with log$ft \approx 6$ and has a much shorter half-life $T_{1/2} \approx 3.7$ hours (see Fig. \ref{fig:decay_scheme}). 

These properties lead to potential applications of $^{176}$Lu in understanding the evolution of various planets and the stellar nucleosynthesis. On one hand, $^{176}$Lu can be used as a nuclear cosmochronometer, which is a powerful tool to study the formation and evolution of planetary bodies in the solar system \cite{Vervoort_Nature_1996}. For example, the Lu-Hf chronometer has been used for the study of crust-mantle evolution of various planetary bodies such as Earth, Moon, Mars, Vesta etc. \cite{PNAS_2015, Bouvier_Nature_2018}. 

On the other hand, in stellar environments, with increasing temperature, the low-lying isomer of $^{176}$Lu can be thermally populated with increasing probability, so that the total stellar $\beta^-$ decay rate of $^{176}$Lu is very sensitive to stellar temperature. This indicates that $^{176}$Lu can be a useful $s$-process thermometer as it is a $s$-only-process nuclide (see Fig. \ref{fig:nuclear_chart}) \cite{Laeter_PRC_2006}. Besides, $^{176}$Lu is one of the important $s$-process branching points \cite{diehl2018astrophysics}.

Although the terrestrial decay rate of $^{176}$Lu has been widely studied experimentally by different methods \cite{Scherer_Science_2001, Amelin_Science_2005, Review_Lithos_2017, 176Lu_Exp_paper_PRC_2023}, the reason that why these (non-unique) first-forbidden transitions are so strange is actually unexplored. This is probably due to the fact that it is challenging for nuclear models to describe forbidden transitions of heavy deformed nuclei (such as the $^{176}$Lu in the rare-earth mass region), which demands large model and configuration spaces as well as reliable many-body wave functions in the laboratory frame with good angular momentum (spin) and parity. Recently, we developed the extended projected shell model (PSM) with large model and configuration spaces where higher-order multi-quasiparticle (qp) configurations are taken into account \cite{LJWang_2014_PRC_Rapid, LJWang_2016_PRC, LJWang_2018_PRC_GT}. Nuclear many-body wave functions are written in the laboratory frame with good spin and parity by the exact projection techniques, and the PSM is further developed to describe allowed \cite{LJWang_2018_PRC_GT, ZRChen_PLB2024} and first-forbidden transitions \cite{BLWang_1stF_2024} of nuclear $\beta$ decays very recently, with the help of the modern Pfaffian algorithm \cite{ZRChen_2022_PRC}. The PSM can adopt deformed (Nilsson) single-particle mean fields to construct configurations, and is then suitable and reliable for description of heavy deformed nuclei. The PSM has been applied successfully in calculating stellar nuclear weak rates \cite{LJWang_PLB_2020_ec, Xiao_Wang_PRC_2024, LJWang_2021_PRL, QYHu_2025_PRL} and bound-state $\beta$ decays \cite{Xiao_PRC_2024_bound_state} as well. Therefore, it should be instructive to try to analyze the underlying mechanisms that derive the anomaly in first-forbidden transitions of $^{176}$Lu by the PSM.


For discussions we show as follows the theoretical framework for the calculation of the terrestrial partial decay rate of first-forbidden transitions \cite{Behrens_NPA_1971, Suhonen_book, Suzuki_2012_PRC, Zhi_FF_PRC_2013, Marketin_QRPA_FF_PRC_2016, BLWang_1stF_2024}, 
\begin{align} \label{eq.lambda_if}
  \lambda^{\beta^-}_{if} 
  = \frac{\ln 2}{K_0} \int_{1}^{Q_{if}} C(W) F_0(Z+1, W) pW (Q_{if} - W)^{2} dW,
\end{align}
from initial ($i$) state of parent nucleus with spin-parity $J_i^{\pi_i}$ to final ($f$) state of daughter nucleus with $J_f^{\pi_f}$. The constant $K_0$ can be determined from superallowed Fermi transitions and $K_0 = 6144 \pm 2$ s \cite{Hardy_2009_PRC} is adopted in this work. $W$ and $p = \sqrt{W^2 - 1}$ label the total energy (rest mass and kinetic energy) and the momentum of the electron in units of $m_e c^2$ and $m_e c$ respectively. The available total energy for leptons in the individual transition is given by $Q_{if} = (M_p - M_d + E_i -E_f ) / m_e c^2$, 
where $M_p$ ($M_d$) indicates the nuclear mass of parent (daughter) nucleus and $E_i$ ($E_f$) labels the excitation energy of the initial (final) state. $F_0$ is the Fermi function.

For non-unique first-forbidden transitions with $|\Delta J| = |J_i-J_f| = 0, 1$, $\Delta \pi = \pi_i \pi_f = -1$, and unique first-forbidden transitions with $|\Delta J|=2$, $\Delta\pi=-1$, the shape factor $C(W)$ has explicit energy dependence, which is approximated as \cite{Weidenmuller_FF_RMP_1961, Behrens_NPA_1971, Zhi_FF_PRC_2013, Mougeot_PRC_2015}, 
\begin{eqnarray} \label{eq.CW}
  C(W) = k(1 + aW + b/W + cW^2) ,
\end{eqnarray}
where the coefficients $k, a, b$ and $c$ depend on the nine reduced nuclear matrix elements in Eq.(\ref{eq.all_ME}) for first-forbidden transition in the following way (see Refs. \cite{Zhi_FF_PRC_2013, BLWang_1stF_2024} for details), 
\begin{eqnarray} \label{eq.kabc}
  k  &=& [\zeta_0^2 + \frac{1}{9} w^2]^{(0)} + [\zeta_1^2 + \frac{1}{9}(x+u)^2 - \frac{4}{9}\mu_1\gamma_1 u(x+u)   \nonumber \\
     & & \quad + \frac{1}{18} Q_{if}^2 (2x+u)^2 - \frac{1}{18} \lambda_2 (2x-u)^2 ]^{(1)} \nonumber \\
     & & \quad + [\frac{1}{12} z^2 (Q_{if}^2-\lambda_2)]^{(2)} , \nonumber \\
  ka &=& [-\frac{4}{3} u Y -\frac{1}{9} Q_{if} (4x^2 +5u^2)]^{(1)} -[\frac{1}{6} z^2 Q_{if}]^{(2)} , \nonumber \\
  kb &=& \frac{2}{3} \mu_1 \gamma_1 \{ -[\zeta_0 w]^{(0)} + [\zeta_1 (x+u)]^{(1)} \} , \nonumber \\
  kc &=& \frac{1}{18} [8u^2 +(2x+u)^2 + \lambda_2 (2x-u)^2]^{(1)} \nonumber \\
     & & \quad +\frac{1}{12} [z^2 (1+\lambda_2)]^{(2)} ,
\end{eqnarray}
where the numbers in parentheses of superscripts denote the rank of the transition operators, $\xi=\alpha Z/(2R)$ with $\alpha$ and $R$ being the fine structure constant and the radius of the nucleus respectively. $\gamma_1=\sqrt{1-(\alpha Z)^2}$ and $\mu_1\approx 1$ are adopted as in Refs. \cite{Zhi_FF_PRC_2013, BLWang_1stF_2024}. $\lambda_2 = \frac{F_1(Z, W)}{F_0(Z, W)}$ with $F_1$ being the generalized Fermi function \cite{Suhonen_PRC_2017_general_Fermi_for_unique}. In Eq. (\ref{eq.kabc}) $\zeta_0, \zeta_1$ and $Y$ are defined as follows, 
\begin{align} \label{eq.VY}
  V &= \xi' \nu + \xi w',   \qquad \ \quad \zeta_0 = V + \frac{1}{3} w Q_{if}, \nonumber \\
  Y &= \xi'y - \xi (u'+x'), \quad  \zeta_1 = Y + \frac{1}{3} (u-x) Q_{if} .  
\end{align}

In Eqs. (\ref{eq.kabc}, \ref{eq.VY}) the involved nuclear matrix elements read as, 
\begin{subequations} \label{eq.all_ME}
\begin{eqnarray}
  w  &=& -g_A \sqrt{3} \frac{\left\langle \Psi^{n_f}_{J_f} \left\| \sum_k r_k [\bm C^k_1 \otimes \hat{\bm\sigma}^k]^0 \hat{\tau}^k_- \right\| \Psi^{n_i}_{J_i} \right\rangle}{\sqrt{2J_i+1}} , \\
  x  &=& - \frac{\left\langle \Psi^{n_f}_{J_f} \left \| \sum_k r_k \bm C^k_1 \hat{\tau}^k_- \right \| \Psi^{n_i}_{J_i} \right \rangle}{\sqrt{2J_i+1}} , \\
  u  &=& -g_A \sqrt{2} \frac{\left\langle \Psi^{n_f}_{J_f} \left \| \sum_k r_k [\bm C^k_1 \otimes \hat{\bm\sigma}^k]^1 \hat\tau^k_- \right \| \Psi^{n_i}_{J_i} \right \rangle}{\sqrt{2J_i+1}} , \\
  z  &=& 2g_A \frac{\left\langle \Psi^{n_f}_{J_f} \left \| \sum_k r_k [\bm C^k_1 \otimes \hat{\bm\sigma}^k]^2 \hat\tau^k_- \right \| \Psi^{n_i}_{J_i} \right \rangle}{\sqrt{2J_i+1}} , \\ 
  w' &=& -g_A \sqrt{3} \frac{\left\langle \Psi^{n_f}_{J_f} \left \| \sum_k \frac{2}{3}r_k I(r_k) [\bm C^k_1 \otimes \hat{\bm\sigma}^k]^0 \hat\tau^k_- \right \| \Psi^{n_i}_{J_i} \right \rangle}{\sqrt{2J_i+1}} , \nonumber \\ \\
  x' &=& - \frac{\left\langle \Psi^{n_f}_{J_f} \left \| \sum_k \frac{2}{3}r_k I(r_k)  \bm C^k_1 \hat\tau^k_- \right \| \Psi^{n_i}_{J_i} \right \rangle}{\sqrt{2J_i+1}} , \\
  u' &=& -g_A \sqrt{2} \frac{\left\langle \Psi^{n_f}_{J_f} \left \| \sum_k  \frac{2}{3}r_k I(r_k) [\bm C^k_1 \otimes \hat{\bm\sigma}^k]^1 \hat\tau^k_- \right \| \Psi^{n_i}_{J_i} \right \rangle}{\sqrt{2J_i+1}} , \nonumber \\ \\
  \xi'\nu &=& \frac{g_A\sqrt{3}}{M_0} \frac{\left\langle \Psi^{n_f}_{J_f} \left \| \sum_k  [\hat{\bm\sigma}_k \otimes \bm\nabla^k]^0 \hat\tau^k_- \right \| \Psi^{n_i}_{J_i} \right \rangle}{\sqrt{2J_i+1}} , \\
  \xi' y  &=& - \frac{1}{M_0} \frac{\left\langle \Psi^{n_f}_{J_f} \big\| \sum_k  \bm\nabla^k \hat\tau^k_- \big\| \Psi^{n_i}_{J_i} \right \rangle}{\sqrt{2J_i+1}}  .
\end{eqnarray} 
\end{subequations}
where the definitions of $\bm C_{lm}$, the radial function $I(r)$ etc. can be seen from Refs. \cite{Zhi_FF_PRC_2013, BLWang_1stF_2024}.


In this work the nuclear matrix elements in Eq. (\ref{eq.all_ME}) are calculated by the PSM where the nuclear many-body wave functions, i.e., the $n$-th eigen state with spin $J$, are expressed in the laboratory frame as \cite{BLWang_1stF_2024}, 
\begin{eqnarray} \label{eq.wave_function}
  | \Psi^{n}_{JM} \rangle = \sum_{K\kappa} f_{K\kappa}^{Jn} \hat{P}_{MK}^{J} | \Phi_{\kappa} \rangle ,
\end{eqnarray}
where $f$ labels the expansion coefficients in the projected basis with $\hat{P}_{MK}^{J}$ being the angular-momentum-projection operator as,
\begin{eqnarray} \label{AMP_operator}
    \hat{P}^{J}_{MK} = \frac{2J + 1}{8\pi^2} \int d\Omega D^{J\ast}_{MK} (\Omega) \hat{R} (\Omega) ,
\end{eqnarray}
where $\hat{R}$ and $D_{MK}^{J}$ (with Euler angle $\Omega$) \cite{varshalovich1988quantum} are the rotation operator and Wigner $D$-function \cite{BLWang_2022_PRC} respectively. $K$ ($M$) is the  projection of the angular momentum in the intrinsic (laboratory) system. In Eq. (\ref{eq.wave_function}) $|\Phi_{\kappa}\rangle$ is the many-body multi-qp configurations which read as \cite{LJWang_2014_PRC_Rapid}, 
\begin{align} \label{eq.config}
  \textrm{oo}: \big\{ & \hat{a}^\dag_{n_i} \hat{a}^\dag_{p_j}|\Phi \rangle, 
               \hat{a}^\dag_{n_i} \hat{a}^\dag_{n_j} \hat{a}^\dag_{n_k} \hat{a}^\dag_{p_l} |\Phi \rangle,
               \hat{a}^\dag_{n_i} \hat{a}^\dag_{p_j} \hat{a}^\dag_{p_k} \hat{a}^\dag_{p_l} |\Phi \rangle,    \big\}  \nonumber \\
  \textrm{ee}: \big\{ & |\Phi \rangle, 
               \hat{a}^\dag_{n_i} \hat{a}^\dag_{n_j} |\Phi \rangle,
               \hat{a}^\dag_{p_i} \hat{a}^\dag_{p_j} |\Phi \rangle,
               \hat{a}^\dag_{n_i} \hat{a}^\dag_{n_j} \hat{a}^\dag_{p_k} \hat{a}^\dag_{p_l} |\Phi \rangle, \nonumber\\ 
             & \hat{a}^\dag_{n_i} \hat{a}^\dag_{n_j} \hat{a}^\dag_{n_k} \hat{a}^\dag_{n_l} |\Phi \rangle, 
               \hat{a}^\dag_{p_i} \hat{a}^\dag_{p_j} \hat{a}^\dag_{p_k} \hat{a}^\dag_{p_l} |\Phi \rangle,   \big\} 
\end{align}
for odd-odd (oo) and even-even (ee) nuclei respectively, where $|\Phi \rangle$ is the qp vacuum with associated intrinsic deformation and $\hat{a}^\dag_n (\hat{a}^\dag_p)$ labels the neutron (proton) qp creation operator.


Figure \ref{fig:decay_scheme} shows the calculated low-lying states and log$ft$ values for the $\beta^-$ decay from $^{176}$Lu to $^{176}$Hf supposing axial symmetry for the parent and daughter nuclei. In the calculations of the log$ft$ values, phenomenological quenching factors are usually introduced artificially in Eq. (\ref{eq.all_ME}) to improve the discrepancy between calculations and data \cite{Suzuki_2012_PRC, Zhi_FF_PRC_2013, Marketin_QRPA_FF_PRC_2016, BLWang_1stF_2024}, it is noted that in this work we focus on the mechanism that leads to the two extremely weak first-forbidden transitions from the $7^-$ state and we avoid adopting any phenomenological quenching factors for the four related first-forbidden transitions shown in Fig. \ref{fig:decay_scheme}. It is seen from Fig. \ref{fig:decay_scheme} that the ground-state band of the daughter nucleus $^{176}$Hf is described well, and the low-lying behavior of the $7^-$ and $1^-$ states of $^{176}$Lu is described reasonably, although the $1^-$ state is calculated to be the ground state. The reason is that the excitation energies of very low-lying states for heavy-deformed odd-odd nuclei are very sensitive to the single-particle energies, pairing correlations, Fermi surfaces etc. for both neutrons and protons. The main configurations are found to be $\nu 7/2^-[514] \otimes \pi 7/2^+[404]$ originating from $\nu 1 h_{9/2}$ and $\pi 1 g_{7/2}$ orbitals respectively with $K^\pi=7^-$ for the $7^-$ state, and $\nu 7/2^-[514] \otimes \pi 5/2^+[402]$ originating from $\nu 1 h_{9/2}$ and $\pi 2 d_{5/2}$ orbitals respectively with $K^\pi=1^-$ for the $1^-$ state. 

It is seen from Fig. \ref{fig:decay_scheme} that when we adopt the small configuration space where only up to 2-qp configurations in Eq. (\ref{eq.config}) are considered, the calculations can well describe the two first-forbidden transitions from the $1^-$ state with the calculated log$ft \approx 6.5$. Besides, the two extremely weak first-forbidden transitions from the $7^-$ state are described reasonably by the calculations with log$ft \approx 17$. When we adopt the large configuration space where a large number of 4-qp configurations in Eq. (\ref{eq.config}) are further considered, the effect on the transitions from the $1^-$ state and the $7^- \rightarrow 8^+$ transition is slight, while the calculated log$ft$ for the $7^- \rightarrow 6^+$ transition is improved noticeably. This indicates that configuration mixing may be important for the two extremely weak first-forbidden transitions from the $7^-$ state.

\begin{table}
  \caption{The calculated dimensionless contributions of vector and tensor terms in the coefficient $k$ in Eq. (\ref{eq.kabc}) for the four related first-forbidden transitions.  } 
  \label{tab1}
\begin{ruledtabular}
\begin{tabular}{ccc}
    Transition & Vector & Tensor \\ \hline
    $1^- \rightarrow 0^+$ & $1.062 \times 10^{-3}$  &  0   \\
    $1^- \rightarrow 2^+$ & $2.031 \times 10^{-3}$  &  $2.386 \times 10^{-9}$    \\ \hline
    $7^- \rightarrow 6^+$ & $9.895 \times 10^{-17}$ &  $3.965 \times 10^{-16}$   \\
    $7^- \rightarrow 8^+$ & $1.810 \times 10^{-14}$ &  $2.857 \times 10^{-16}$ 
\end{tabular}
\end{ruledtabular}
\end{table}

The transition strength of first-forbidden transition is determined by the nine nuclear matrix elements in Eq. (\ref{eq.all_ME}), which are divided into the scalar–axial ($w$ and $w'$), vector–vector ($x$ and $x'$), vector–axial ($u$ and $u'$) and tensor–axial ($z$) terms, as well as the terms associated with the relativistic corrections ($\xi'\nu$ and $\xi' y$) \cite{Behrens_NPA_1971, Suhonen_book}. In particular, the terrestrial partial decay rate or log$ft$ is determined by the coefficient $k$ in Eq. (\ref{eq.kabc}) which includes scalar, vector and tensor components. To analyze the mechanism for the two anomalous first-forbidden transitions from the $7^-$ state, we first show in Table \ref{tab1} the vector and tensor components in the coefficient $k$ in Eq. (\ref{eq.kabc}) for the four related first-forbidden transitions. The scalar components vanish because the spins of initial and final states are different from each other for all the related transitions. It is seen that for the transitions from the $1^-$ state, the vector component dominates the the coefficient $k$, while for the transitions from the $7^-$ state, both vector and tensor components contribute significantly. Actually we found from the PSM calculations that for the two first-forbidden transitions from the $7^-$ state, the tensor term $z$ in Eq. (\ref{eq.all_ME}d) turns out to be the largest in the nine nuclear matrix elements in Eq. (\ref{eq.all_ME}).

\begin{table}
  \caption{The log$ft$ value, the configuration mixing percentage, and the average $K$ value $\bar K$, from the PSM calculations with different configuration spaces (no mixing case, small configuration space and large configuration space, respectively) for the $7^- \rightarrow 6^+ \ (8^+)$ first-forbidden transition of $^{176}$Lu. See the text for details.  } 
  \label{tab2}
\begin{ruledtabular}
\begin{tabular}{cccc}
  Transition & \multicolumn{3}{c}{$7^- \longrightarrow 6^+ \ (8^+)$} \\ \cline{2-4} 
                                  &     No mixing    &     Small               &    Large                 \\ \hline
  \multirow{2}{*}{log$ft$}        &     19.95        &    17.82                &   19.42                  \\ 
                                  &     (21.69)      &    (17.29)              &   (17.49)                \\ \hline  
  \multirow{2}{*}{Config-mixing}  &  0               &  $3.96\%$               &  $39.16\%$               \\
                                  &  0 (0)           &  $55.11\%$  ($64.00\%$) &  $66.36\%$  ($72.96\%$)  \\ \hline  
  \multirow{2}{*}{ $\bar K$ }     &  7               &  6.824                  &  6.809                   \\
                                  &  0 (0)           &  0.197 (0.320)          &  0.252 (0.276)           \\  
\end{tabular}
\end{ruledtabular}
\end{table}

It is well known that there exist many multi-qp high-$K$ isomers in the rare-earth region \cite{Walker_1999_Nature, Walker_2001_Hyperfine, Sun_2005_Nature_Phys}. The nuclear isomers in $^{176}$Hf have been describe successfully by our PSM calculations with extended configuration space \cite{LJWang_2014_PRC_Rapid}. The mechanism of the multi-qp high-$K$ isomers can be understood as follows. For electromagnetic transitions with $|K_i - K_f| > \lambda$ where $K_i$ ($K_f$) denotes the $K$ value of the initial (final) state and $\lambda$ labels the rank of the transition operators, the corresponding intrinsic matrix elements of the electromagnetic transition operators vanish since it is not possible to conserve the $K$ quantum number for nuclei with axial symmetry \cite{Bohr_Nuclear_Structure_book_1998} in the intrinsic system. Transitions of this type are referred to as $K$ forbidden and $\rho \equiv |K_i - K_f| - \lambda$ is introduced to denote the order of $K$ forbiddenness \cite{Bohr_Nuclear_Structure_book_1998}. However, since $K$ is not an observable and the nuclear wave functions have to be written in the laboratory system instead of the intrinsic system, $K$-forbidden transitions with small $\lambda$ (for example, the $E2$ transitions) can still occur, the corresponding transition is usually weak and the transition strength is determined in a very sensitive way by the beyond-mean-field effect, the configuration mixing and $K$ mixing during the transformation from the intrinsic to laboratory system. To some extent, it is expected that the $K$-forbidden transitions with larger $\lambda$ (i.e., smaller $\rho$) tend to be stronger. Similar to the above electromagnetic transition, the $K$-forbidden mechanism may be important for first-forbidden transition during nuclear $\beta$ decays as well. As seen from Fig. \ref{fig:decay_scheme} and the corresponding discussions, for the $7^- \rightarrow 6^+ (8^+)$ first-forbidden transition of $^{176}$Lu, one may expect that the initial $7^-$ state has $K^\pi \approx 7^-$, and the final $6^+$ and $8^+$ states have $K^\pi \approx 0^+$ (because the $6^+$ and $8^+$ states belong to the ground-state rotational band which is based on the qp vacuum $|\Phi \rangle$). Therefore, for these two $K$-forbidden $\beta^-$-decay first-forbidden transitions, the nuclear matrix elements of the tensor term $z$ in Eq. (\ref{eq.all_ME}d), which have the smallest $\rho$ with $\rho \equiv |K_i - K_f| - \lambda = 7-2=5$ in all the scalar, vector and tensor terms in Eq. (\ref{eq.all_ME}), should be the largest. This is consistent with the conclusion shown in Table \ref{tab1} and the corresponding discussions.

To analyze the underlying mechanisms that lead to the anomaly in the first-forbidden transitions of $^{176}$Lu, we show in Table \ref{tab2} the log$ft$ values for the $7^- \rightarrow 6^+ \ (8^+)$ transition calculated by the PSM with different configuration spaces. The no mixing case corresponds to keeping only the 2-qp configuration $\nu 7/2^-[514] \otimes \pi 7/2^+[404]$ with $K^\pi=7^-$ for the $7^-$ state for $^{176}$Lu, and only the qp vacuum $|\Phi \rangle$ with $K^\pi=0^+$ for $^{176}$Hf, (which is the main configuration in realistic calculation), neglecting any possible configuration mixing. The small configuration space corresponds to only up to 2-qp configurations in Eq. (\ref{eq.config}) are considered, and the large configuration space corresponds to that a large number of 4-qp configurations in Eq. (\ref{eq.config}) are further considered. The configuration mixing percentage and the average $K$ value $\bar K$ of the $7^-$ state (upper) and the $6^+$ and $8^+$ states (lower) are listed as well. The configuration mixing percentage denotes the mixing percentage of all other multi-qp configurations except for the main configuration discussed above, which can be extracted by the wave function $f$ in the projected basis $\hat{P}_{MK}^{J} | \Phi_{\kappa} \rangle $ which has good $K$ quantum number from Eq. (\ref{eq.wave_function}). $\bar K$ can be obtained in the same way. It is seen that when there is no configuration mixing, one gets log$ft \approx 20$ and 22 for the two transitions, the transitions do not vanish because of the beyond-mean-field effect from the angular-momentum projection. 

When the small configuration space is adopted, the configuration mixing is small for the initial $7^-$ state and very large for the final $6^+$ and $8^+$ states. However, as analyzed from our PSM calculations, the wave function of the initial $7^-$ state only includes mixing of multi-qp configurations with $K^\pi = 6^-$ or $7^-$, and those of the final $6^+$ and $8^+$ states only include mixing of multi-qp configurations with $K^\pi = 0^+$ or $1^+$. This leads to small $K$ mixing with $\bar K = 6.824$, 0.197 and 0.320 for the $7^-$, $6^+$ and $8^+$ states respectively, so that the transitions turn out to be stronger with log$ft = 17.82$ and 17.29, as seen from Table \ref{tab2}. When the large configuration space is adopted further, the configuration mixing becomes very large for initial and final states, while the $K$ mixing is still very small due to the same reason mentioned above in the case of small configuration space. Compared with the case of small configuration space, the $7^- \rightarrow 6^+ \ (8^+)$ transition becomes weaker with log$ft = 19.42$ (17.49), which is due to that the mixed configurations do not contribute coherently. This indicates that the detailed configuration mixing and $K$ mixing are crucial for the transitions of nuclear $\beta$ decays with the possible $K$-forbidden nature.

From our realistic PSM calculations with large configuration space, the calculated half-life of the $7^-$ state is $T_{1/2} \approx 1.95 \times 10^{10}$ years. By comparison, the experimental data range is $T_{1/2} = (1.90-7.50) \times 10^{10}$ years \cite{176Lu_half_life_PRC_2003}. Finally, it is worth mentioning that both the high-$K$ isomers and $K$-forbidden $\beta$-decay transitions provide a good tool to check and constrain nuclear models in the rare-earth region.


In summary, the investigation of mechanisms in nuclear physics are important for understanding many interdisciplinary problems. The two extremely weak first-forbidden transitions from the $7^-$ ground state of $^{176}$Lu is the key for understanding the evolution of planetary bodies in the solar system, the temperature of the $s$ process, the path of the $s$ process in this branching point etc. Based on our projected shell model which is developed very recently to describe first-forbidden transitions of nuclear $\beta$ decay in a realistic way, the mechanism for the anomaly in first-forbidden transitions of $^{176}$Lu is studied for the first time. The $K$-forbidden nature is found to be the key for the anomaly in the first-forbidden transitions, and the extremely weak transition strengths are determined in a sensitive way by the angular-momentum projection which transform the nuclear wave function from the intrinsic to the laboratory system, and by the detailed configuration mixing and $K$ mixing in the nuclear wave functions in the laboratory system. Our discussion is important for understanding the $K$ quantum number and the intrinsic system in quantum (nuclear) many-body problems.

\begin{acknowledgments}
  We thank Y. Xiao for checking the calculations, and F.-Q. Chen for helpful discussions. This work is supported by the National Natural Science Foundation of China (Grant No. 12275225). 
\end{acknowledgments}


%

\end{document}